\documentclass[prb,twocolumn]{revtex4}
\usepackage{amsfonts}
\usepackage{amsmath}
\usepackage{amssymb}
\usepackage{graphicx}%
\begin{document}
\title{First-principles LDA+$U$ and GGA+$U$ study of plutonium oxides}
\author{Bo Sun}
\affiliation{Institute of Applied Physics and Computational Mathematics, P.O. Box 8009,
Beijing 100088, P.R. China}
\author{Ping Zhang}
\thanks{Corresponding author. Electronic address: zhang\_ping@iapcm.ac.cn}
\affiliation{Institute of Applied Physics and Computational
Mathematics, P.O. Box 8009, Beijing 100088, P.R. China}
\author{Xian-Geng Zhao}
\affiliation{Institute of Applied Physics and Computational
Mathematics, P.O. Box 8009, Beijing 100088, P.R. China}

\begin{abstract}
The electronic structure and properties of PuO$_{2}$ and Pu$_{2}$O$_{3}$ have
been studied from first principles by the all-electron
projector-augmented-wave (PAW) method. The local density approximation
(LDA)+$U$ and the generalized gradient approximation (GGA)+$U$ formalism have
been used to account for the strong on-site Coulomb repulsion among the
localized Pu $5f$ electrons. We discuss how the properties of PuO$_{2}$ and
Pu$_{2}$O$_{3}$ are affected by the choice of $U$ as well as the choice of
exchange-correlation potential. Also, oxidation reaction of Pu$_{2}$O$_{3}$,
leading to formation of PuO$_{2}$, and its dependence on $U$ and
exchange-correlation potential have been studied. Our results show that by
choosing an appropriate $U$ it is promising to correctly and consistently
describe structural, electronic, and thermodynamic properties of PuO$_{2}$ and
Pu$_{2}$O$_{3}$, which enables it possible the modeling of redox process
involving Pu-based materials.

\end{abstract}
\maketitle

\section{Introduction}

Plutonium dioxide (PuO$_{2}$) and sesquioxide (Pu$_{2}$O$_{3}$) are the only
observed stoichiometric compounds formed at the surface of the metallic
plutonium when exposed to dry air \cite{Has2000} (nonstoichiometric oxide may
form by reaction of dioxide with water \cite{Has}). From this sense, plutonium
corrosion and oxidation are often treated as equivalent topic. The plutonium
corrosion plays a key role in considering the nuclear stockpile and storage of
surplus plutonium. Therefore, a thorough understanding of the physical and
chemical properties of the plutonium oxide is always needed.

From basic point of view, it can be visualized that many physical and chemical
properties of the plutonium oxide are closely related to the quantum process
of localization and delocalization for Pu 5$f$ electrons. Modeling of the
electron localization, and thus any redox process involving plutonium, is a
complex task. Conventional density functional schemes that apply the local
density approximation (LDA) or the generalized gradient approximation (GGA)
underestimate the strong on-site Coulomb repulsion of the Pu 5$f$ electrons
and consequently fail to capture the correlation-driven localization.
Therefore, the 5$f$ electrons in elemental Pu, as well as in Pu compounds,
require special attention. One promising way to improve contemporary LDA and
GGA approaches is to modify the intra-atomic Coulomb interaction through the
so-called LDA+$U$ or GGA+$U$ approach, in which the underestimation of the
intraband Coulomb interaction is corrected by the Hubbard $U$ parameter
\cite{Ani1991,Ani1993,Sol1994}. This method has been used to discuss the
equilibrium lattice parameter of bulk Pu in Ref. \cite{Sav2000,Shick1,Shick2}.
The choice of $U$ is, however, not unambiguous and it is not trivial to
determine its value \textit{a priori}, though there are attempts to extract it
from standard first-principles calculations. Hence, $U$ is often fitted to
reproduce a certain set of experimental data such as band gaps and structural properties.

In this paper we use the LDA+$U$ and GGA+$U$ schemes due to Dudarev et al.
\cite{Dud} to calculate the lattice parameters, electronic structure, and
thermodynamic properties of PuO$_{2}$ and Pu$_{2}$O$_{3}$. We discuss how
these properties are affected by the choice of $U$ as well as the choice of
exchange-correlation potential, i.e., the LDA or the GGA, and how redox
processes occurred in plutonium oxide can be explored in the LDA+$U$ and
GGA+$U$ formalism. In addition, we notice that recently there have occurred a
few experimental \cite{But2004,But2006,Gou2007} and theoretical
\cite{But2006,Prodan2005,Prodan2006,Prodan2007} studies of the electronic
structures of plutonium oxides. In this paper we have compared our calculated
LDA/GGA+$U$ results with those reports. Our results show that while the pure
LDA/GGA (without $U$ correction) calculations fail to describe the
ground-state behaviors of the plutonium oxides, such as the insulating nature,
the magnetic configuration, and the $5f$ band gap, the present LDA/GGA+$U$
approaches with tunable Coulomb parameters can effectively remedy those
failures and the consequent results fit well in the attainable experimental
data \cite{But2004,But2006,Gou2007}.

This paper is organized as follows. The details of our calculations are
described in Sec. II and in Sec. III we present and discuss the results. In
Sec. IV, we summarize our findings.

\section{Methodology of the calculation}

The calculations were performed using the projector-augmented wave (PAW)
method of Bl\"{o}chl \cite{Blo}, as implemented in the ab initio total-energy
and molecular-dynamics program VASP (Vienna \textit{ab initio} simulation
program) \cite{Kresse1,Kresse2,Kresse3,Kresse4}. PAW is an all-electron method
that combines the accuracy of augmented-plane-wave methods with the efficiency
of the pseudopotential approach. The PAW method is implemented in VASP with
the frozen-core approximation. For the plane-wave set, a cut-off energy of 400
eV was used. The plutonium 6$s$, 6$p$, 7$s$, and 5$f$, and the oxygen 2$s$ and
2$p$ electrons were treated as valence electrons. The strong on-site Coulomb
repulsion amongst the localized Pu $5f$ electrons are accounted for by using
the formalism formulated by Dudarev \textit{et al}. \cite{Dud}. In this scheme
the total LDA (GGA) energy functional is of the form%
\begin{equation}
E_{\text{LDA(GGA)+}U}=E_{\text{LDA(GGA)}}+\frac{U-J}{2}\sum_{\sigma}\left[
\text{Tr}\rho^{\sigma}-\text{Tr}\left(  \rho^{\sigma}\rho^{\sigma}\right)
\right]  ,\label{e1}%
\end{equation}
where $\rho^{\sigma}$ is the density matrix of $f$ states, and $U$ and $J$ are
the spherically averaged screened Coulomb energy and the exchange energy,
respectively. In this paper the Coulomb $U$ is treated as a variable, while
the exchange energy is set to be a constant $J$=0.75 eV. This value of $J$ is
in the ball park of the commonly accepted one for Pu
\cite{Sav2000,Mar1988,Shick1999,Shick2005}. Since only the difference between
$U$ and $J$ is significant \cite{Dud}, thus we will henceforth label them as
one single parameter, for simplicity labeled as $U$, while keeping in mind
that the non-zero $J$ has been used during calculations.

The exchange and correlation effects were treated in both the local density
approximation and the generalized gradient approximation \cite{Perdew}. We
studied PuO$_{2}$ in its ground-state fluorite structure ($Fm3m$) and the
sesquioxide Pu$_{2}$O$_{3}$ in the hexagonal $\beta$--type structure
($P\bar{3}m1$). For PuO$_{2}$ we used a $11\times11\times11$ Monkhorst-Pack
$k$-point mesh \cite{Monk} (56 irreducible $k$ points) and for Pu$_{2}$O$_{3}$
we used a $9\times9\times6$ grid (57 irreducible $k$ points). The electronic
density of states (DOS) was obtained with $15\times15\times15$ (120
irreducible $k$ points) and $11\times11\times9$ grid (120 irreducible $k$
points) $k$-point meshes, respectively. The Brillouin-zone integration was
performed using the modified tetrahedron method of Bl\"{o}chl \cite{Blo2}. In
order to study the reaction energy it is necessary to calculate the energy of
an oxygen molecule ($E_{\text{O}_{2}}$). The effect of spin polarization was
included in calculating $E_{\text{O}_{2}}$.

\section{Results and discussion}

\subsection{Atomic and electronic structure of PuO$_{2}$}%

\begin{figure}[tbp]
\begin{center}
\includegraphics[width=1.0\linewidth]{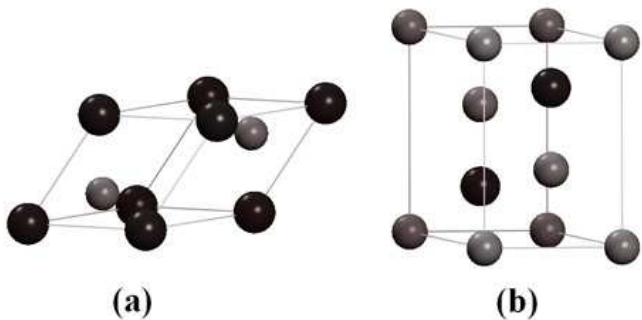}
\end{center}
\caption{(a) Unit cell of PuO$_{2}$ containing 3 atoms. The black
spheres are Pu atoms, the gray ones are oxygens. (b) Unit cell of
Pu$_{2}$O$_{3}$ containing five atoms.}
\label{fig1}
\end{figure}%
Plutonium dioxide crystallizes in a CaF$_{2}$-like ionic structure [Fig. 1(a)]
with the plutonium and oxygen atoms forming face-centered and simple cubic
sublattices, respectively. In this arrangement each plutonium atom is located
at the center of an oxygen cube, and for every four such cubes there is an
empty one. In the ionic limit, formal charge for plutonium in PuO$_{2}$ is +4,
corresponding to formal population of $f^{4}$. This leads to local $S$=2
plutonium moment, which can couple with other sites in either a ferromagnetic
(FM) or antiferromagnetic (AFM) manner. PuO$_{2}$ is known to be an insulator
\cite{Mc1964} and some scattered experimental data \cite{San1999} support the
ground state of PuO$_{2}$ to be an AFM phase. In the present LDA/GGA+$U$
approaches, we have considered the FM, AFM, and nonmagnetic phases for each
choice of the value of $U$ and then determined the ground-state phase by a
subsequent total-energy comparison of these three phases. For PuO$_{2}$, at
$U$=$0$, the ground state is a FM metal, which is in contrast to experiment.
By increasing the amplitude of $U$, our LDA/GGA+$U$ calculations correctly
predicted an AFM insulating ground state. The turning value of $U$ for this
FM-AFM energy transition of the ground state is of $\sim$1.5 eV. In the
discussion that follows, we therefore confined our report to the AFM solution
for the PuO$_{2}$. A thorough description of the magnetic structure of
plutonium oxides is beyond our intention in this paper, and we would like to
leave it for the future studies.

The experimentally determined lattice parameter of PuO$_{2}$ is $a_{0}%
$=$5.396$ \AA at 25$^{\circ}$C \cite{Haire2001}. Here the calculated $a_{0}$
and bulk modulus $B_{0}$ of PuO$_{2}$ were obtained from the corresponding
energy minimization at constant volumes and by fitting a Murnaghan equation of
state \cite{Mur1944} to the resulting energy-volume data, respectively. The
results as a function of $U$ within the LDA and the GGA schemes are shown in
Fig. 2(a) and (b) for $a_{0}$ and $B_{0}$, respectively. For comparison, the
experimental values of $a_{0}$ \cite{Haire2001} and $B_{0}$ \cite{Idi2004} are
also shown in Fig. 2.%
\begin{figure}[tbp]
\begin{center}
\includegraphics[width=1.0\linewidth]{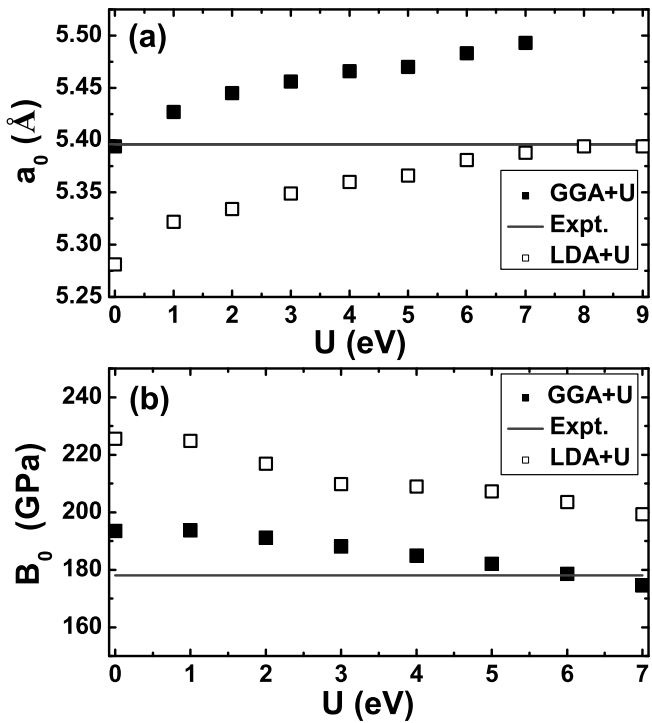}
\end{center}
\caption{Dependence of the lattice parameter (a) and bulk modulus
(b) of PuO$_{2}$ on $U$.} \label{fig2}
\end{figure}
For the pure DFT calculation ($U$=$0$), it shows in Fig. 2(a) that the LDA
overbinds the compound and underestimates with respect to the experiment the
lattice parameter by $\sim$2\%, while the GGA calculation give a slight
overestimate of $a_{0}$. After turning on the Hubbard $U$, one can see from
Fig. 2(a) that for the LDA+$U$ approach, although the lattice parameter is
still underestimated in a wide range of $U$, the calculated $a_{0}$ for
PuO$_{2}$ improves upon the pure LDA by steadily increasing its amplitude with
$U$. In fact, at a typical value of $U$=$4$ eV, the LDA+$U$ gives $a_{0}$=5.36
\AA , which is very close to the experiment. On the other hand, with
increasing $U$, the underbind effect brought about by the GGA+$U$ is somewhat
enlarged. As a comparison, at $U$=$4$ eV, the GGA+$U$ gives $a_{0}$=5.47 \AA ,
which overestimates the experimental data by $\sim$1.3\%. Overall, both the
LDA+$U$ and GGA+$U$ results of the lattice parameter for the PuO$_{2}$ AFM
phase are comparable with experiment by tuning in the calculations the Hubbard
$U$ around $4$ eV. We have also calculated the equilibrium lattice parameter
for the FM and nonmagnetic phases for PuO$_{2}$. The tendency of $a_{0}$ with
$U$ for these two phases is similar to that for the present AFM phase. For the
calculated bulk modulus $B_{0}$ of the PuO$_{2}$ AFM phase, one can see from
Fig. 2(b) that its value varies with $U$ over a rather broad range of 175 to
195 GPa for the GGA+$U$ and 200 to 230 GPa for the LDA+$U$. The LDA result of
$B_{0}$ is always higher than the GGA result, which is due to the
above-mentioned \textquotedblleft overbind\textquotedblright\ characteristics
of the LDA approach. For the measurements of the equilibrium bulk modulus,
there are no consistent results to date for the AFM PuO$_{2}$. Here we
compared our calculation to the experimental result of $B_{0}$=$178$ GPa
reported in Ref. \cite{Idi2004}. One can see from Fig. 2(b) that the
discrepancy between the present calculation and the experiment is most
distinct at $U$=$0$. Both the LDA and the GGA give an overestimate, with the
latter more close to the experimental data. By turning on the effective
Coulomb interaction, the amplitude of $B_{0}$ begins to decrease. At a typical
value of $U$=$4$ eV, the LDA+$U$ gives $B_{0}$=208 GPa while the GGA+$U$ gives
$B_{0}$=184 GPa. We notice that the recent hybrid density-funtional
calculations \cite{Prodan2006} predict the bulk modulus to be 220 GPa for the
antiferromagnetic PuO$_{2}$, comparable with the present pure LDA results. For
lattice parameter $a_{0}$ it was predicted to be 5.46 in Ref.
\cite{Prodan2006}. To conclude (Fig. 2), comparing with the experimental data
and the recent hybrid density-functional results, the accuracy of our
atomic-structure prediction for the antiferromagnetic PuO$_{2}$ is quite
satisfactory by tuning the effective Hubbard parameter $U$ in a range $3$--$4$
eV within the LDA/GGA+U approaches.%

\begin{figure}[tbp]
\begin{center}
\includegraphics[width=1.0\linewidth]{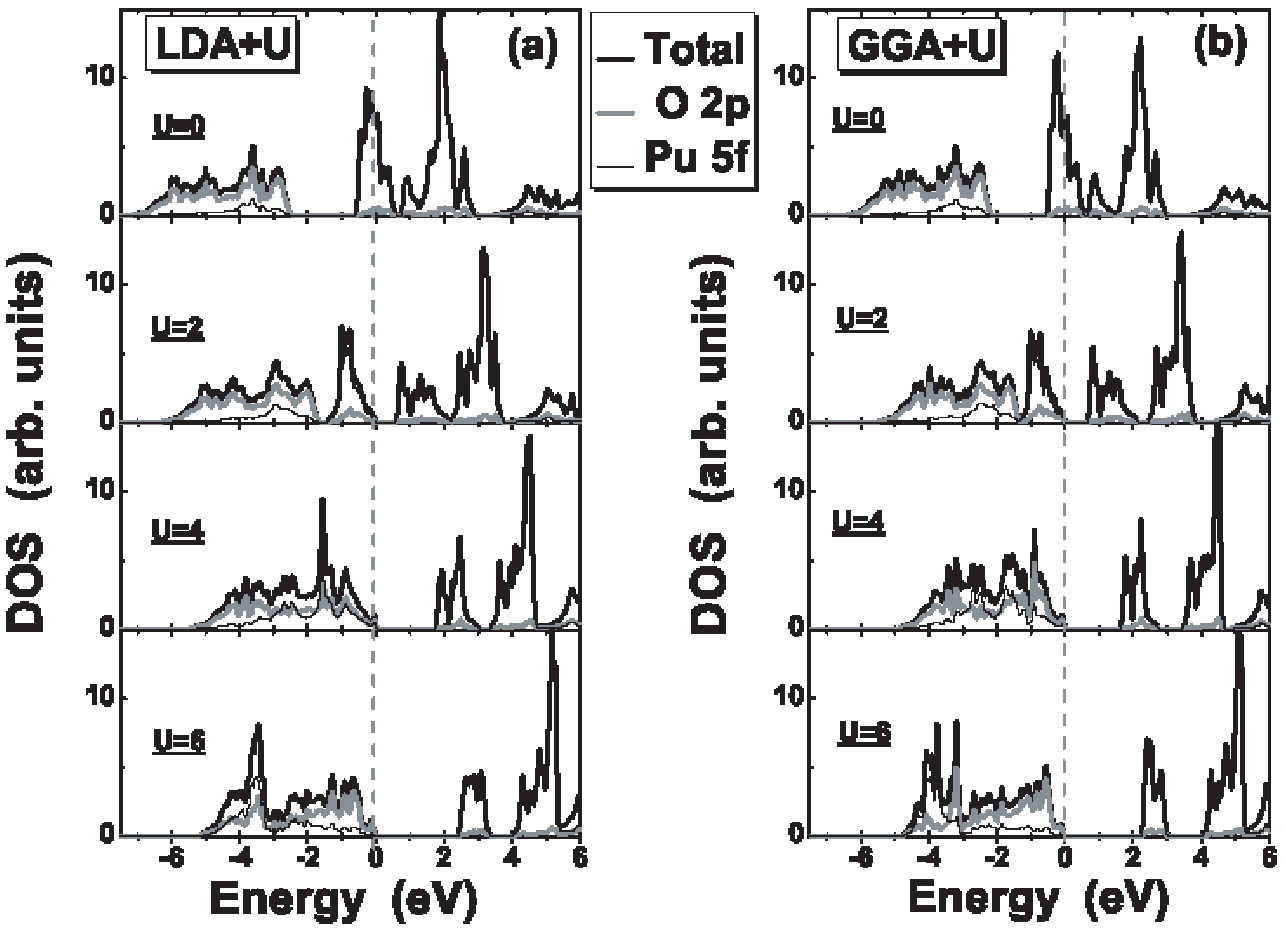}
\end{center}
\caption{The total DOS for the PuO$_{2}$ antiferromagnetic phase
computed in the (a) LDA+$U$ and (b) GGA+$U$ formalism with four
selective values of $U$. The projected DOS for the Pu $5f$ and O
$2p$ orbitals are also shown. The Fermi level was set to be zero.}
\label{fig3}
\end{figure}%
Besides the prominent changes in the atomic-structure parameters, the most
dramatic improvement brought by the LDA/GGA+$U$ when compared to the pure ones
is in the description of electronic-structure properties. For this we have
investigated the band structures of the PuO$_{2}$ AFM phase with the aim at
seeing the fundamental influence by the inclusion of the on-site Coulomb
interaction. The resultant total density of states (DOS) for four selective
values of $U$ are plotted in left (LDA+$U$) and right (GGA+$U$) panels in Fig.
3. For more clear illustration, the projected DOS for the Pu $5f$ and O $2p$
orbitals are also shown in Fig. 3.\ The Fermi energy $E_{F}$ has been set to
be zero. Without accounting for the on-site Coulomb repulsion ($U$=$0$), one
can see that both two pure DFT methods predict an incorrect metallic ground
state by non-zero occupation of Pu 5$f$ states at $E_{F}$. When switching on
$U$, as shown in Fig. 3, the Pu 5$f$ band begins to split at $E_{F}$ and tends
to open a gap $\Delta$. The amplitude of this insulating gap increases with
increasing $U$, see Fig. 4.%
\begin{figure}[tbp]
\begin{center}
\includegraphics[width=1.0\linewidth]{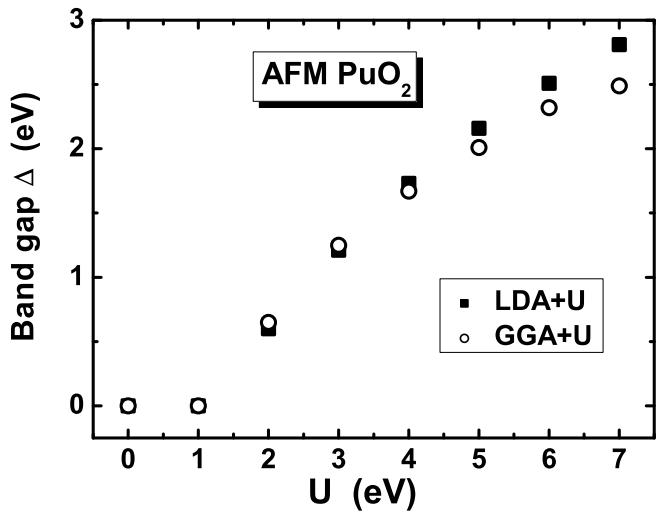}
\end{center}
\caption{The insulating band gap of the PuO$_{2}$ antiferromagnetic
phase as a function of $U$ for the LDA (filled squares) and the GGA
(hollow circles).} \label{fig4}
\end{figure}
Overall the LDA+$U$ and GGA+$U$ give an equivalent description of the
one-electron behaviors in a wide range of $U$. At a typical value of $U$=4 eV,
one can see from Fig. 3 that the occupied DOS is featured by two well-resolved
peaks. The narrow one near $-$2.0 eV is principally Pu $5f$ in character,
while the broad one near $-$4.0 eV is mostly O 2$p$. These two pronounced
peaks have been observed in the recent photoemission measurements
\cite{But2004,Gou2007}. In addition, by increasing the amplitude of $U$, one
prominent feature occurred in Fig. 3 is the increasing hybridization between
Pu $5f$ and O $2p$ occupied states. This interesting mixing effect disappears
in the cases of Pu$_{2}$O$_{3}$ (see Fig. 6 below) and UO$_{2}$
\cite{Kudin2002}, for which the Pu (U) $5f$ and O $2p$ occupied bands are well
separated. The presence of Pu($5f$)-Pu($2p$) hybridization in PuO$_{2}$
implies a more covalent and stronger metal-ligand mixing than in Pu$_{2}%
$O$_{3}$ and UO$_{2}$. This phenomenon appears surprising, given the smaller
overlap anticipated in Pu because of the smaller radius of the Pu $5f$
orbital. Experimentally, Butterfield \textit{et al}. \cite{But2004} and Gouder
\textit{et al}. \cite{Gou2007} have reported the thin-film photoemission data
for PuO$_{2}$. The present overall picture which emerges from the LDA/GGA+$U$
with properly selective Coulomb repulsion appear to be in satisfactory
agreement with experiment. We have also compared our results given in Fig. 3
with the most recent calculations by Prodan \textit{et al}. \cite{Prodan2006}
based on newly developed screened Coulomb hybrid density functional. The
agreement between our LDA/GGA+$U$ (with $U\sim$4 eV) results and those in Ref.
\cite{Prodan2006} is also apparent. Interestingly, the above-mentioned orbital
(Pu $5f$ and O $2p$) mixing effect in PuO$_{2}$ has also been theoretically
predicted by Prodan \textit{et al}. \cite{Prodan2006,Prodan2007}, who
hypotheses that the expected stabilization of the Pu $5f$ orbital energy
relative to U $5f$ leads to an \textquotedblleft accidental\textquotedblright%
\ degeneracy between the Pu $5f$ and O $2p$ levels, which in the first-order
perturbation theory results in a higher degree of covalency regardless of
small radius of the Pu $5f$ orbital. Therefore, although the pure LDA and GGA
fail to depict the electronic structure, especially the insulating nature and
the occupied-state character of PuO$_{2}$, our present results show that by
tuning the effective Hubbard parameter in a reasonable range, the LDA/GGA+$U$
approaches will prominently improve upon the pure LDA/GGA calculations and
thus can provide a satisfactory qualitative electronic structure description
comparable with experiments and the hybrid DFT calculation. By further
increasing $U$ to 6 eV, one can see that the peak near $-$2.0 eV becomes weak
and is mostly O $2p$, while the peak near $-$4.0 eV becomes stronger and
consists equally of Pu 5$f$ and O 2$p$ orbital. This picture of DOS is no
longer valid since the peak near $-$2.0 eV has been confirmed to be due to the
Pu 5$f$ contribution. Thus the LDA/GGA+U approaches with $U$ as large as 6 eV
fails to describe the electronic structure of PuO$_{2}$.

\subsection{Atomic and electronic structure of Pu$_{2}$O$_{3}$}

Pu$_{2}$O$_{3}$ is an insulating oxide of the hexagonal $\beta$-type
($P\bar{3}m1$) [Fig. 1(b)] with space group no. 164, the only phase of the
sesquioxide that has been prepared with stoichiometric composition. Both
magnetic susceptibility \cite{Mc1981} and neutron diffraction \cite{Wul1988}
measurements have found Pu$_{2}$O$_{3}$ to have an AFM structure at
temperatures below 4.2 K, with the Pu moments $\mu$ confined along the
$\mathbf{z}$ axis in a simple +$-$+$-$ alternation of spins. As with PuO$_{2}%
$, we have considered the FM, AFM, and nonmagnetic phases and then determined
the ground-state phase by comparing the equilibrium total energies of these
three phases. At $U$=$0$, the calculated ground state is as for PuO$_{2}$ an
incorrect FM metal. By increasing the amplitude of $U$, our LDA/GGA+$U$
approaches correctly predicted the $\beta$--Pu$_{2}$O$_{3}$ to be in an AFM
insulating phase. The FM-AFM energy crossing occurs at a small $U$ of $\sim
$1.5 eV. We report in what follows on the Pu$_{2}$O$_{3}$ AFM phase.%

\begin{figure}[tbp]
\begin{center}
\includegraphics[width=1.0\linewidth]{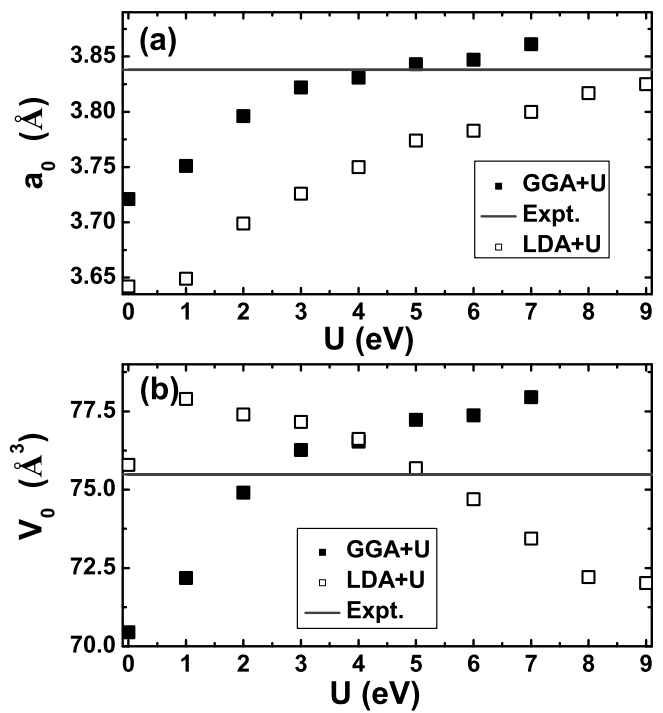}
\end{center}
\caption{Dependence of the equilibrium lattice parameter $a_{0}$ (a)
and the volume $V_{0}$ of unit cell (b) of Pu$_{2}$O$_{3}$ on $U$.}
\label{fig5}
\end{figure}%
The calculated equilibrium lattice parameter $a_{0}$ of Pu$_{2}$O$_{3}$ is
plotted in Fig. 5(a) as a function of $U$. It reveals that the relation
between $a_{0}$ and $U$ does not follow a simple monotonic function. The
turning point is at $U$=$3$ eV, below which $a_{0}$ goes up rapidly with $U$.
After crossing this turning point, the increase of $a_{0}$ begins to slow down
with $U$. Thus the curvature of $a_{0}$ for small values of $U$ is more
significant than for large values of $U$. The decrease in curvature at large
$U$ corresponds to the separation of the occupied Pu $5f$ band from the
unoccupied part, i.e., the transition from a metallic to an insulating ground
state of Pu$_{2}$O$_{3}$ (see below). This feature in the increase of $a_{0}$
as a function of $U$ is almost the same for the LDA and the GGA. The
experimental data \cite{Mc1981,Wul1988} of $a_{0}$=3.841 \AA  is well fitted
at $U$=4 eV for the GGA, while the LDA always slightly underestimates $a_{0}$.
Another feature shown in Fig. 5(a) is that at small values of $U$ below 4 eV
the GGA underestimates $a_{0}$, which is contrary to the general experience
that in most cases (as shown for PuO$_{2}$), the GGA often gives a slight
overestimate of lattice parameter. This rarely-occurred feature may be due to
the appearance of the other lattice parameter in $\beta$--Pu$_{2}$O$_{3}$,
i.e., the rario $c_{0}/a_{0}$ for the hexagonal crystalline structure. The
equilibrium volume $V_{0}$ of of the Pu$_{2}$O$_{3}$ unit cell (including 5
atoms) as a function of $U$ is plotted in Fig. 5(b). The experimental result
\cite{Mc1981,Wul1988} of $V_{0}$ is also given for comparison. Although the
tendency of $V_{0}$ with $U$ is remarkably opposite for the two DFT+$U$
methods, the results mostly overlap at a typical value of $U$=$4$ eV, at which
insulating gap for the Pu$_{2}$O$_{3}$ is well formed. The different tendency
of $V_{0}$ with respect to $U$ for the LDA and GGA may come from sensitivity
of the anisotropy in Pu $5f$ orbitals to the treatment of the
exchange-correlation potential. Combining Fig. 5(a) and (b) it is expected
that both the LDA and the GGA may give a satisfactory prediction of the
ground-state atomic structure for the Pu$_{2}$O$_{3}$ by tuning $U$ to be near
$4$ eV.

The LDA/GGA+$U$ total DOS for the Pu$_{2}$O$_{3}$ AFM phase are shown in Fig.
6 for four selective values of $U$.%
\begin{figure}[tbp]
\begin{center}
\includegraphics[width=1.0\linewidth]{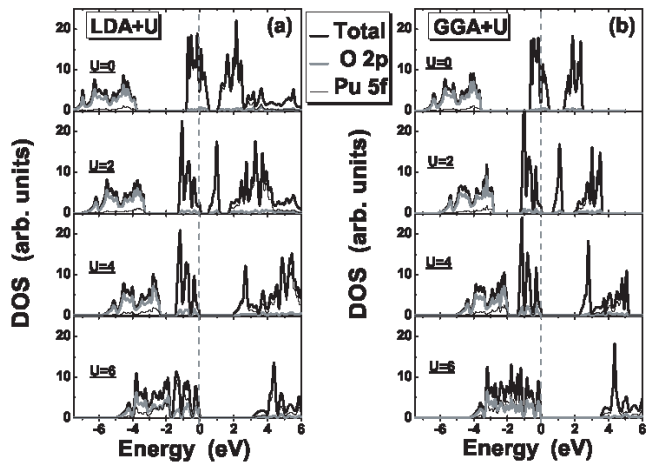}
\end{center}
\caption{The total DOS for the Pu$_{2}$O$_{3}$ antiferromagnetic
phase computed in the (a) LDA+$U$ and (b) GGA+$U$ formalism with
four selective values of $U$. The projected DOS for the Pu $5f$ and
O $2p$ orbitals are also shown.}
\label{fig6}
\end{figure}
The projected DOS for the Pu $5f$ and O $2p$ orbitals are also plotted.\ Both
the LDA and GGA predict an incorrect metallic ground state for Pu$_{2}$O$_{3}$
at $U$=$0$ by the presence of non-zero occupation of Pu 5$f$ state at the
Fermi energy $E_{F}$. When turning on the on-site Coulomb repulsion, the Pu
5$f$ band begins to split and form an insulating gap $\Delta$ at a critical
value $U$=$1$ eV. The gap $\Delta$ becomes large with increasing $U$, as shown
in Fig. 7, from which one can see that the amplitude of $\Delta$ for Pu$_{2}%
$O$_{3}$ is almost equivalent to that for PuO$_{2}$ at low $U$. At a typical
value of $U$=4 eV, it reveals in Fig. 6 that the occupied DOS is featured by
two peaks. The narrow one near $-$1.5 eV is principally Pu $5f$ in character,
while the broad one around $-$4.0 eV is mostly O 2$p$. It is encouraging that
these two pronounced peaks, as well as the overall appearance of the total DOS
spectrum, fit well in recent photoemission experiments \cite{But2004,Gou2007}
on Pu$_{2}$O$_{3}$. We have also compared our results given in Fig. 6 with the
recent calculations by Prodan \textit{et al}. \cite{Prodan2006} using the
hybrid density functional. Our LDA/GGA+$U$ results (with $U\sim$4 eV) for the
Pu$_{2}$O$_{3}$ AFM phase are in excellent agreement with those in Ref.
\cite{Prodan2006}. Unlike in PuO$_{2}$, the Pu $5f$ and O $2p$ states in
Pu$_{2}$O$_{3}$ are well separated in the DOS spectrum. This feature is
similar to that of UO$_{2}$ \cite{Kudin2002}, which also exhibits two distinct
peaks of U $5f$ and O $2p$ parentage. Remarkably, the similar trend has also
been theoretically reported on Pu$_{2}$O$_{3}$ in Ref. \cite{Prodan2006}
within the hybrid-density-functional framework. A consistent explanation with
the Pu($5f$)-O($2p$) hybridization in PuO$_{2}$ may sustain by understanding
the orbital separation in Pu$_{2}$O$_{3}$ as a consequence of the more weakly
bound Pu $5f$ site energy associated with the less highly charged Pu$^{3+}$
ion \cite{Prodan2005}. With further increasing the effective intratomic
Coulomb interaction to $U$=$6$ eV, as shown in Fig. 6, the separation of the
Pu $5f$ from O $2p$ projected DOS is blurred by the increasing spectrum weight
of the former around $-4$ eV, which overlaps largely with the O $2p$. This no
longer accord with the experiments \cite{But2004,Gou2007}. Therefore, as with
PuO$_{2}$, the LDA/GGA+U approaches with $U$ as large as 6 eV fail to describe
the electronic structure of Pu$_{2}$O$_{3}$.%
\begin{figure}[tbp]
\begin{center}
\includegraphics[width=1.0\linewidth]{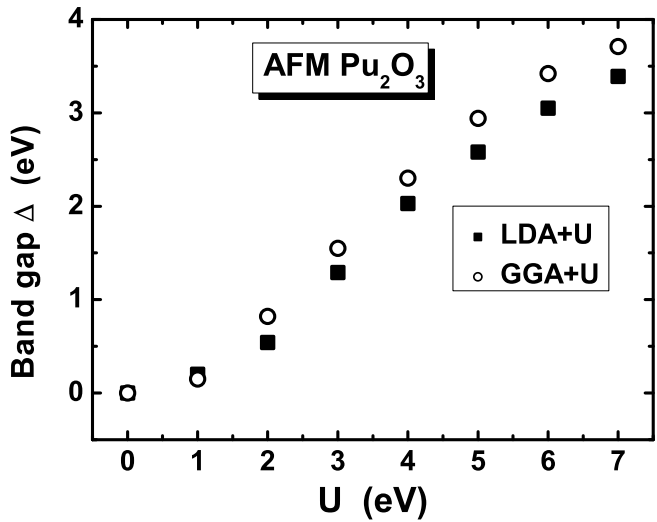}
\end{center}
\caption{The insulating band gap of the Pu$_{2}$O$_{3}$
antiferromagnetic phase as a function of $U$ for the LDA (filled
squares) and the GGA (hollow circles).}
\label{fig7}
\end{figure}%

\subsection{Oxidation reaction energy}

Oxidation of Pu$_{2}$O$_{3}$ via the reaction
\begin{equation}
\text{Pu}_{2}\text{O}_{3}+\frac{1}{2}\text{O}_{2}\rightarrow2\text{PuO}_{2}
\label{e2}%
\end{equation}
leads to formation of stoichiometric PuO$_{2}$. The dependence of the
transformation reaction energy on $U$ is presented in Fig. 8.
\begin{figure}[tbp]
\begin{center}
\includegraphics[width=1.0\linewidth]{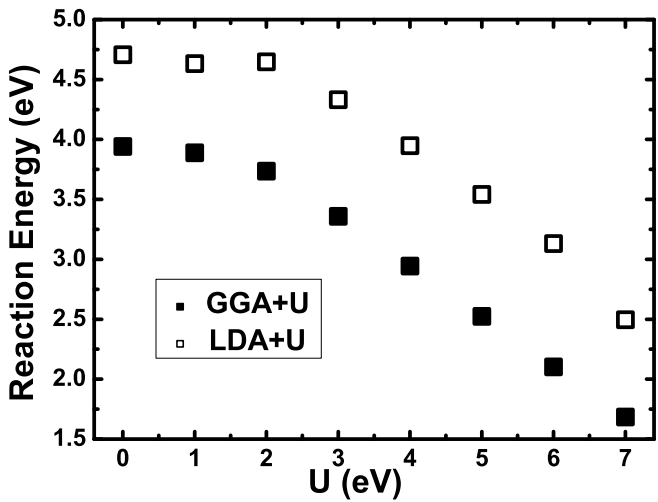}
\end{center}
\caption{Dependence of the
Pu$_{2}$O$_{3}+\frac{1}{2}$O$_{2}\rightarrow 2$PuO$_{2}$ reaction
energy on $U$.}
\label{fig8}
\end{figure}
One can see that both the LDA and the GGA show the same dependence of reaction
energy on the on-site Coulomb interaction. That is, at small values of $U$
which correspond to the metallic ground state for both PuO$_{\text{2}}$ and
Pu$_{2}$O$_{3}$, the reaction energy is independent of $U$. Above the
metallic-insulating transition, our calculated reaction energy decreases
linearly with increasing $U$. The reason for this behavior is that a high $U$
favors localization and thus facilitates the transition. Density functional
theory is known to overestimate the binding energy of O$_{2}$ and this should
result in an underestimation of the present reaction energy via the
$E_{\text{O}_{2}}$ term. Consequently we cannot expect a perfect agreement
with experiments for the present reaction energy. However, this error is
independent of any conditions in the plutonium oxide and thus can be remedied
by shifting the energy of O$_{2}$ so as to give the experimental binding
energy. In the LDA the O$_{2}$ binding energy is overestimated by 1.2 eV/0.5
O$_{2}$ and in the GGA the corresponding number is 0.8 eV/0.5 O$_{2}$. The GGA
always predicts a lower value of the reaction energy, as seen from Fig. 9.

\section{Conclusions}

We have studied the structural, electronic, and thermodynamic properties of
the antiferromagnetic PuO$_{2}$ and Pu$_{2}$O$_{3}$ within the LDA+$U$ and
GGA+$U$ frameworks. The atomic structure, including lattice parameters and
bulk modulus, and the one-electron behaviors of these kinds of plutonium
oxides have been systematically investigated as a function of the effective
on-site Coulomb repulsion parameter $U$. We find that both the LDA+$U$ and
GGA+$U$ considerably improves upon the traditional density functionals,
providing a first-principles description of plutonium oxides in satisfactory
qualitative agreement with experiment. Also our present results are well
comparable to those obtained through newly developed hybrid DFT method.
Specially, from the LDA/GGA+$U$ study of the lattice parameter of PuO$_{2}$ we
find that the experimental data of $a_{0}$ can be gradually approached by
steadly increasing $U$ to be in an acceptable range around 4 eV. The incorrect
metallic ground state at purely LDA or GGA ($U$=$0$) for both PuO$_{2}$ and
Pu$_{2}$O$_{3}$ can be readily corrected by a systematic inclusion of non-zero
$U$, which forces the Pu 5$f$ band to split at the Fermi level and thus drives
the metallic-insulating transition. The insulating band gaps for PuO$_{2}$ and
Pu$_{2}$O$_{3}$ have been shown as a function of $U$. The oxidation reaction
Pu$_{2}$O$_{3}$+0.5O$_{2}\rightarrow$2PuO$_{\text{2}}$ has also been studied
by systematically calculating the reaction energy as a function of $U$. Our
results show that the oxidation process of the Pu$_{2}$O$_{3}$ is an
exothermic reaction, which is mostly responsible for the experimentally
observed \cite{Martz1994} plutonium pyrophoricity at 150$^{\text{o}}$%
C$-$200$^{\text{o}}$C. Also we have shown that above the metallic-insulating
transition, the reaction energy decreases with increasing $U$ for the LDA and
the GGA schemes. We expect these calculated results are useful for the future
studies on the surface oxidation and corrosion of metallic plutonium.

\begin{acknowledgments}
This work was partially supported by NSFC under grants Nos. 10604010 and 60776063.
\end{acknowledgments}


\begin{thebibliography}{99}                                                                                               %


\bibitem {Has2000}J.M. Haschke, Los Alamos Science \textbf{26}, 253 (2000).

\bibitem {Has}J.M. Haschke, T.H. Allen, and L.A. Morales, Science
\textbf{287}, 285 (2000).

\bibitem {Ani1991}V.I. Anisimov, J. Zaanen, and O.K. Anderson, Phys. Rev. B
\textbf{44}, 943 (1991).

\bibitem {Ani1993}V.I. Anisimov, I.V. Solovyev, M.A. Korotin, M.T. Czy\.{z}yk,
and G.A. Sawatzky, Phys. Rev. B \textbf{48}, 16929 (1993).

\bibitem {Sol1994}I.V. Solovyev, P.H. Dederichs, and V.I. Anisimov, Phys. Rev.
B \textbf{50}, 16861 (1994).

\bibitem {Sav2000}S.Y. Savrasov and G. Kotliar, Phys. Rev. Lett. \textbf{84},
3670 (2000).

\bibitem {Shick1}A.B. Shick, V. Drchal, and L. Havela, Europhys. Lett.
\textbf{69}, 588 (2005).

\bibitem {Shick2}A. Shick, L. Havela, J. Koloren\v{c}, V. Drchal, T. Gouder,
and P.M. Oppeneer, Phys. Rev. B \textbf{73}, 104415 (2006).

\bibitem {Dud}S.L. Dudarev, G.A. Botton, S.Y. Savrasov, C.J. Humphreys, and
A.P. Sutton, Phys. Rev. B \textbf{57}, 1505 (1998).

\bibitem {But2004}M. Butterfield, T. Durakiewicz, E. Guziewicz, J. Joyce, A.
Arko, K. Graham, D. Moore, and L. Morales, Surf. Sci. \textbf{571}, 74 (2004).

\bibitem {But2006}M.T. Butterfield, T. Durakiewicz, I.D. Prodan, G.E.
Scuseria, E. Guziewicz, J.A. Sordo, K.N. Kudin, R.L. Martin, J.J. Joyce, A.J.
Arko, K.S. Graham, D.P. Moore, and L.A. Morales, Surf. Sci. 600, 1637 (2006).

\bibitem {Gou2007}T. Gouder, A. Seibert, L. Havela, and J. Rebizant, Surf.
Sci. 601, L77 (2007).

\bibitem {Prodan2005}I.D. Prodan, G.E. Scuseria, J.A. Sordo, K.N. Kudin, and
R.L. Martin, J. Chem. Phys. \textbf{123}, 014703 (2005).

\bibitem {Prodan2006}I.D. Prodan, G.E. Scuseria, and R.L. Martin, Phys. Rev. B
\textbf{73}, 045104 (2006).

\bibitem {Prodan2007}I.D. Prodan, G.E. Scuseria, and R.L. Martin, Phys. Rev. B
\textbf{76}, 033101 (2007).

\bibitem {Blo}P.E. Bl\"{o}chl, Phys. Rev. B \textbf{50}, 17953 (1994).

\bibitem {Kresse1}G. Kresse and J. Hafner, Phys. Rev. B \textbf{48}, 13115 (1993).

\bibitem {Kresse2}G. Kresse and J. Furthm\"{u}ller, Comput. Mater. Sci.
\textbf{6}, 15 (1996).

\bibitem {Kresse3}G. Kresse and J. Furthm\"{u}ller, Phys. Rev. B \textbf{54},
11169 (1996).

\bibitem {Kresse4}G. Kresse and D. Joubert, Phys. Rev. B \textbf{59}, 1758 (1999).

\bibitem {Mar1988}D. van der Marel and G.A. Sawatzky, Phys. Rev. B
\textbf{37}, 10674 (1988); J.F. Herbst, R.E. Watson, and I. Lindgren,
\textit{ibid}. \textbf{14}, 3265 (1976).

\bibitem {Shick1999}A.B. Shick, A. I. Liechtenstein, and W.E. Pickett, Phys.
Rev. B \textbf{60}, 10763 (1999).

\bibitem {Shick2005}A.B. Shick, V. Jani\v{s}, and P.M. Oppeneer, Phys. Rev.
Lett. \textbf{94}, 016401 (2005).

\bibitem {Perdew}J.P. Perdew, J.A. Chevary, S.H. Vosko, K.A. Jackson, M.R.
Pederson, D.J. Singh, and C. Fiolhais, Phys. Rev. B \textbf{46}, 6671 (1992).

\bibitem {Monk}H.J. Monkhorst and J.D. Pack, Phys. Rev. B \textbf{13}, 5188 (1976).

\bibitem {Blo2}P.E. Bl\"{o}chl, O. Jepsen, and O.K. Andersen, Phys. Rev. B
\textbf{49}, 16223 (1994).

\bibitem {Mc1964}C. E. McNeilly, J. Nucl. Mater. \textbf{11}, 53 (1964).

\bibitem {San1999}P. Santini, R. L\'{e}manski, and P. Erd\~{o}s, Adv. Phys.
\textbf{48}, 537 (1999); M. Colarieti-Tosti, O. Eriksson, L. Nordstr\"{o}m, J.
Wills, and M.S.S. Brooks, Phys. Rev. B \textbf{65}, 195102 (2002); S. Kern, R.
A. Robinson, H. Nakotte, G. H. Lander, B. Cort, P. Watson, and F. A. Vigil,
\textit{ibid}. \textbf{59}, 104 (1999); G. Raphael and R. Lallement, Solid
State Commun. \textbf{6}, 383 (1968).

\bibitem {Haire2001}R. G. Haire, J. M. Haschke, MRSBull. 689 (September 2001).

\bibitem {Mur1944}F.D. Murnaghan, Proc. Natl. Acad. Sci. U.S.A. \textbf{30},
244 (1944).

\bibitem {Idi2004}M. Idiri, T. LeBihan, S. Heathman, and J. Rebizant, Phys.
Rev. B \textbf{70}, 014113 (2004).

\bibitem {Kudin2002}K.N. Kudin, G.E. Scuseria, and R.L. Martin, Phys. Rev.
Lett. \textbf{89}, 266402 (2002).

\bibitem {Mc1981}B. McCart, G.H. Lander, and A.T. Aldred, J. Chem. Phys.
\textbf{74}, 5263 (1981).

\bibitem {Wul1988}M. Wulff and G.H. Lander, J. Chem. Phys. \textbf{89}, 3295 (1988).

\bibitem {Ell1961}F.H. Ellinger, The Metal Plutonium (The University of
Chicago Press, Chicago, IL, 1961).

\bibitem {Martz1994}J.C. Martz, J.M. Haschke, and J.L. Stakebake, J. Nuclear
Materials \textbf{210}, 130 (1994).
\end{thebibliography}
\end{document}